\documentclass[psamsfonts]{amsart}

\usepackage{amssymb,amsfonts}
\usepackage[all,arc]{xy}
\usepackage{enumerate}
\usepackage{mathrsfs}
\usepackage[toc,page]{appendix}
\usepackage[top=2cm, bottom=2cm, left=3cm, right=3cm]{geometry}
\usepackage{multicol}
\usepackage{graphicx}
\RequirePackage{color,graphicx}
\usepackage{hyperref}
\definecolor{linkcolour}{rgb}{0,0.2,0.6}
\hypersetup{colorlinks,breaklinks,urlcolor=linkcolour, linkcolor=linkcolour}
\usepackage[utf8]{inputenc}
\usepackage[anythingbreaks]{breakurl}

\usepackage{tabularx}

\theoremstyle{definition}

\theoremstyle{remark}

\makeatletter
\let\c@equation\c@thm
\makeatother
\numberwithin{equation}{section}

\title{\textsc{D\MakeLowercase{emonstrating research subcommunities in mathematical networks}}}

\author{Steven B. Bradlow, Konstantinos Kapenekakis, Georgios Kydonakis, Xinwei Li and Jiarui Xu}

\begin{document}
\maketitle

\begin{abstract}
We propose a method for demonstrating sub community structure in scientific networks of relatively small size from analyzing databases of publications. Research relationships between the network members can be visualized as a graph with vertices corresponding to authors and with edges indicating joint authorship. Using a fast clustering algorithm combined with a graph layout algorithm, we demonstrate how to display these clustering results in an attractive and informative way. The small size of the graph allows us to develop tools that keep track of how these research sub communities evolve in time, as well as to present the research articles that create the links between the network members. These tools are included in a web app, where the visitor can easily identify the various sub communities, providing also valuable information for administrational purposes. Our method was developed for the GEAR mathematical network and it can be applied to other networks.

\end{abstract}
\begin{multicols}{2}
\section{Introduction}

Bibliometric mapping provides an effective tool to quickly summarize and then visualize research relationships among the members of scientific networks. The resulting visualizations can also identify potential future research directions and collaboration opportunities (see [1]). In this paper we describe how we applied these ideas to one particular network, namely the GEAR Network (described below), but the methods clearly have wide applicability. Among the various possible types of bibliometric networks and visualization approaches (see [5]), we chose to build a co-authorship network visualized as a graph, with vertices corresponding to authors and weighted edges indicating joint authorship (see [4]).

The GEAR Network was created in 2011 through the U.S. National Science Foundation's Research in Mathematical Sciences (RNMS) program. The acronym GEAR stands for ``GEometric structures And Representation varieties'' and reflects the mathematical focus of the network.  The GEAR network brings together over 350 researchers from 90 nodes throughout the world, aiming `to promote research interactions between its members and facilitate the cross-pollination of ideas between different research subgroups encompassed by the network'.

The starting point in order to build a graph demonstrating research sub communities within the GEAR network was to create a database including joint authored papers by the Network members, published in mathematical journals or being in the preprint status. For our purposes we included only articles authored by at least two GEAR members that have appeared since the establishment of the Network in 2011. We then converted this information into a graph with vertices representing authors and weighted edges indicating joint authorship, where this `weight' describes the number of publications for a particular pair of co-authors. The weights are a key feature that the algorithms we employ are using in order to detect sub community structure in the graph.

In order to detect and visualize the sub community structure in our Network, we used the Gephi network visualization software. Gephi can translate joint authorship data into a highly readable graph, which includes deep-level information. This information can be extracted in a format that is convenient for visual display.

In this direction, our team also built an interactive interface to display the cluster results obtained using the tools described.  We have created a web app where the visitor can easily identify the various research sub communities within the GEAR network. Nodes and lines in this version of our graph are made clickable, providing information about the corresponding GEAR members, their collaborators within the Network along with the jointly-authored papers, while a sidebar allows the user to employ a node-locator tool.

This interactive interface allows one to understand the nature of the network as a whole and observe how the collaborations among its members evolve in time. This can be particularly useful for the network's administration, as it can point out suggestions for organizing future research meetings in order to strengthen relationships among different research sub communities.

\section{Building a graph from the database of jointly authored papers}

The database including jointly authored papers was built by processing publication information for the Network members.

The next step was to convert this into a graph demonstrating collaborations, with vertices corresponding to authors and edges corresponding to jointly authored papers. We created in .csv format a matrix describing the number of papers between each pair of GEAR Members that appear as co-authors in our papers database. This number of papers served as ``weight'' for every edge and we will see below how the clustering algorithm that we employed, is using this weight to detect community structure in the graph. The matrix constructed provided an appropriate way to input the publication information collected into the very powerful visualization tool of the Gephi software, which is commonly used for creating and displaying clustering results for similar graphs.

\section{Clustering algorithms and visualization}

Gephi is an open-source network visualization software package written in Java, and is ideal in displaying the spatialization process. We used Gephi to apply a clustering algorithm for our graph, a graph layout algorithm and then we exported this coordinate and clustering information. The input for Gephi was a .csv file which included id numbers of the network members and the number of jointly authored papers for every collaboration taking place within our network.

Running the modularity function in Gephi applies a fast unfolding algorithm by V. Blondel, J. Guillaume, R. Lambiotte and E. Lefebvre [2] for detecting community structure of large networks. The modularity of a partition of a network into communities is a scalar value in the interval $\left[ -1,1 \right]$ that measures the density of links inside communities, as opposed to links between communities. Thus, networks with high modularity indicate dense connections among the nodes within clusters, but sparse connections with nodes from different clusters. For weighted networks (like ours, considering a weight for each edge to represent the number of papers written by the two nodes), the modularity is defined as \[Q=\frac{1}{2m}\sum\limits_{i,j}{\left[ {{A}_{ij}}-\frac{{{k}_{i}}{{k}_{j}}}{2m} \right]}\,\delta \left( {{c}_{i}},{{c}_{j}} \right)\]
where ${{A}_{ij}}$ represents the weight of the edge between $i$ and $j$, ${{k}_{i}}=\sum\nolimits_{j}{{{A}_{ij}}}$ is the sum of the weights of the edges attached to vertex $i$, ${{c}_{i}}$ is the community to which vertex $i$ is assigned, the $\delta $-function $\delta \left( u,v \right)$ is $1$ if $u=v$ and $0$ otherwise, and $m=\frac{1}{2}\sum\limits_{ij}{{{A}_{ij}}}$.

The algorithm is composed of two phases repeated iteratively. During the first phase, each node is treated as being a separate cluster. For each node, the gain of modularity by removing the node from its cluster and placing it in a neighbor's cluster is computed. Then the node is placed in the cluster for which the gain is maximum, and the process is repeated until no further improvement can be achieved. In the second phase of the algorithm, a new network is being built, whose nodes are now the clusters found during the first phase. As soon as this process terminates, community structure is marked by changing the color of the nodes according to their modularity class. The software provides several functions to improve the graph's appearance by adjusting the color, size and shape of the graph.

Moreover, Gephi provides multiple layout algorithms to exhibit the graph. We chose the ForceAtlas2 algorithm [3] to do this. This algorithm simulates a physical system in order to spatialize a network. The nodes are regarded as charged particles and the edges attract the nodes like springs. The positions of the nodes and edges are adjusted depending on the forces of attraction and repulsion until these eventually converge to a balanced state. The attraction force ${{F}_{a}}$ between two connected nodes ${{n}_{1}}$ and ${{n}_{2}}$ depends linearly on their distance $d\left( {{n}_{1}},{{n}_{2}} \right)$. The repulsion force ${{F}_{r}}$ is proportional to the product of the degrees plus one of the two nodes
	\[{{F}_{r}}\left( {{n}_{1}},{{n}_{2}} \right)={{k}_{r}}\frac{\left[ \deg ({{n}_{1}})+1 \right]\left[ \deg ({{n}_{2}})+1 \right]}{d({{n}_{1}},{{n}_{2}})}\]
where by degree of a node we mean the number of edges meeting at this node and the coefficient ${{k}_{r}}$ is defined from the settings.

Eventually, in order to display the results on a web app and moreover include more detailed information on the graph's nodes and edges (such as author names, list of collaborators and publication information for the joint articles), we extracted this information in a .json format file, which is a highly readable format and can be easily imputed into other interactive tools for creating a web app.

\section{Displaying the results on a webpage}

The ``GEAR Collaboration Graph'' \href{http://gear.math.illinois.edu/members/signin/collaboration/}{\nolinkurl{[http://gear.math.illinois.edu/members/signin/collaboration/]}} is an online app written in the web technologies html, css and javascript.

\begin{center}
  \includegraphics[scale=0.3]{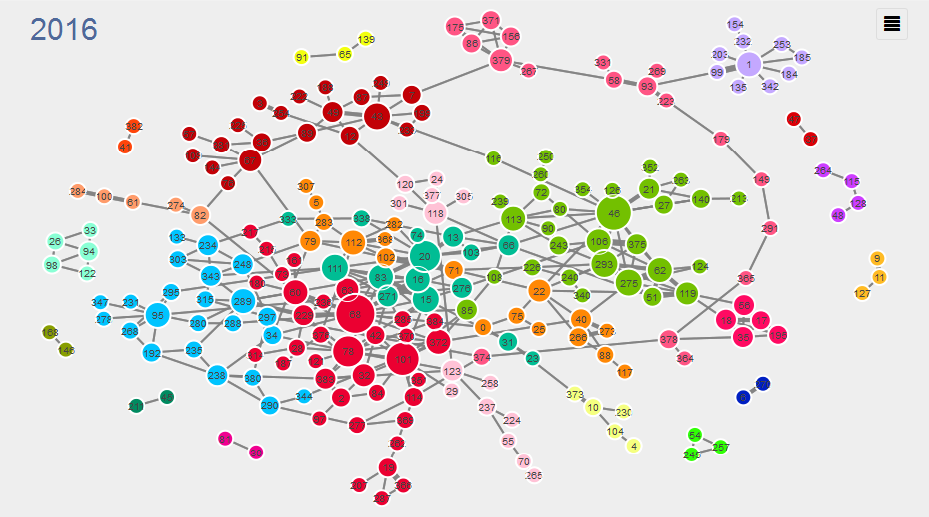}
\end{center}

We used the twitter bootstrap \href{http://getbootstrap.com/}{\nolinkurl{(http://getbootstrap.com/)}} framework for page skeleton and styling, which takes care of common basic web development problems and speeds up the development process. For the rest styling features we used plain css code written using BEM \href{https://en.bem.info/}{\nolinkurl{(https://en.bem.info/)}}, which is a method of naming css classes and is ideal for making future development improvements.

Page functionality and graph representation was built using javascript. We used the jQuery library \href{https://jquery.com/}{\nolinkurl{(https://jquery.com/)}}, which facilitates manipulating the web page’s elements, as well as underscore \href{http://underscorejs.org/}{\nolinkurl{(http://underscorejs.org/)}}, which provides essential tools for data objects and arrays manipulation such as sorting and parsing items.

We also used the handlebars templating library \href{http://handlebarsjs.com/}{\nolinkurl{(http://handlebarsjs.com/)}}, which abstracts the html code building in javascript.

Therefore, the main process is fetching the data files which are in json format, parsing member items, sorting the items in groups that represent collaborations, and placing items as svg nodes in the graph. Then we are connecting associated graph items with clickable lines that represent members' joint research work.

Moreover, we wanted to include data from several different years, in order to exhibit the evolution in time of our graph, which would increase the time a page viewer is waiting for each particular graph to be built. At this point, the external app of Gephi, which is written in java, was used for the graph design. Gephi precalculates the positions of the graph's nodes using more sophisticated mathematical functions that would be too heavy for a web browser to execute.

\begin{center}
\includegraphics[scale=0.14]{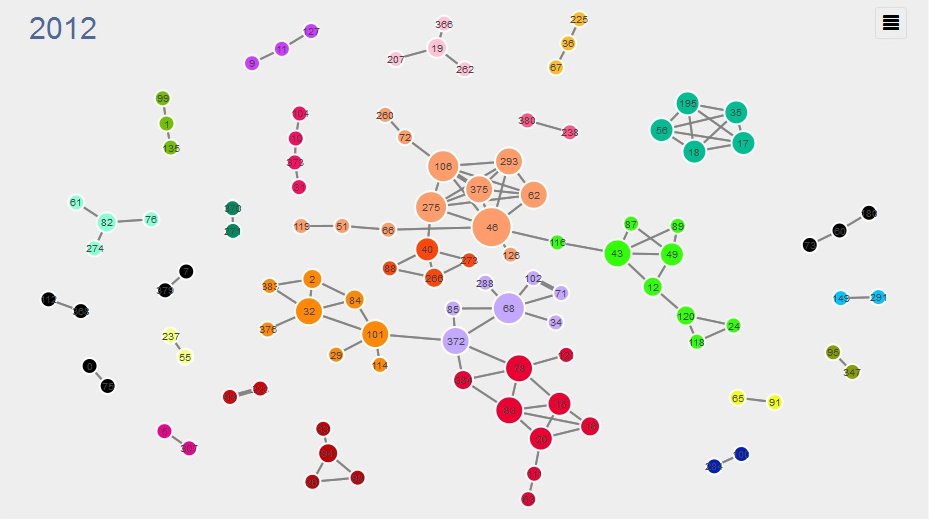} \includegraphics[scale=0.14]{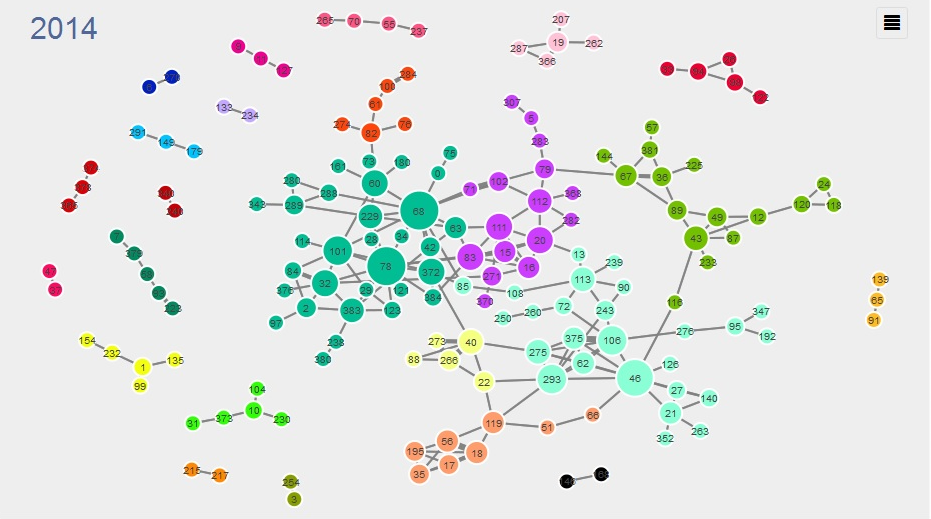}
\end{center}

The graph drawing functionality was built using d3 \href{https://d3js.org/}{\nolinkurl{(https://d3js.org/)}}, which provides us with fast mathematical functions and algorithms designed to quickly draw scalable graphics for the web. We also used the Q library \href{https://github.com/kriskowal/q/}{\nolinkurl{(https://github.com/kriskowal/q/)}}, which helps running many asynchronous tasks and makes the drawing process even faster.

Apart from the ‘evolution-in-time’ feature, the nodes and edges of the graph were made clickable, providing information about the GEAR collaborators of a particular GEAR member as well as their jointly-authored papers. Finally, a sidebar allows the user to employ a node-locator tool.

 \begin{center}
  \includegraphics[scale=0.25]{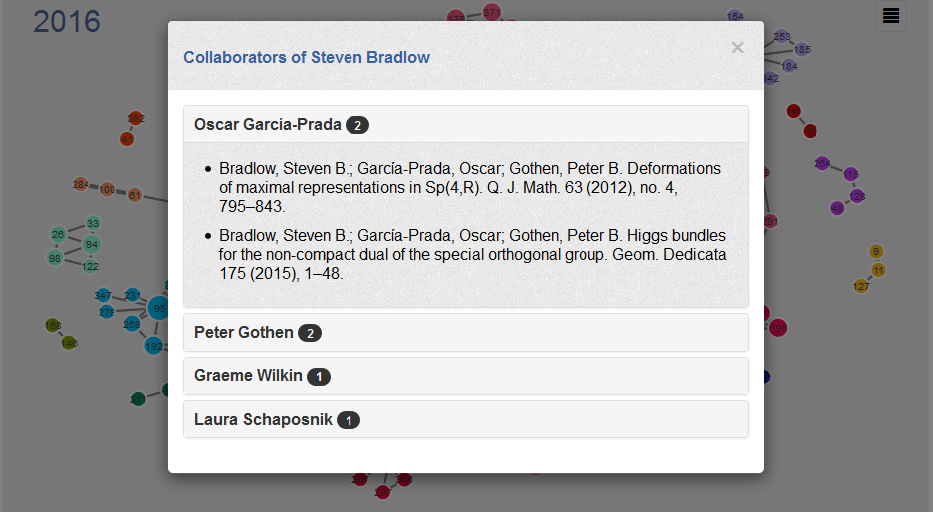}
\end{center}

\section{Features revealed and further applications}

The clustering and graph-drawing algorithms described here are very appropriate in detecting sub community structure for networks of relatively small size (and not just for such). The results are being displayed in a way that combined with the online tools included, provide us with valuable information about the nature of the network as a whole, and observe the evolution in time of the collaborations between its members. Table 1 below summarizes interesting statistics about the collaborations among the GEAR Network members that has resulted in the publication of a research article from 2011 until September 2016.

A visitor to the GEAR website can easily realize the research interactions taking place among the Network's members. The research sub communities apparent within the Network are directly exhibited, along with specific reference to the particular work leading to the formulation of these sub communities (joint papers published).

However, such a graph can be also useful for administrational purposes. It may provide a means of measuring how much a network is achieving its scope in promoting research interactions between members, who may belong in somewhat different research groups. Moreover, focusing on what appear to be the ``boundaries'' between different research sub communities, can point out suggestions for organizing future research events to further explore connections between areas in mathematics that originally were not thought to be so strongly related.
\end{multicols}

\begin{table}
  \caption{GEAR Collaboration Graph Statistics}
  \begin{tabularx}{\textwidth}{XXXXXX}
   Years included & Co-authors (nodes) & Connected  components & Clusters  & Mean distance between nodes & Modularity \\
    \hline\hline
    2011-2011   &  66  & 22 & 22 & 1.49 & 0.905 \\
    2011-2012   & 106  & 23 & 27 & 3.41 & 0.879 \\
    2011-2013   & 131  & 20 & 24 & 4.04 & 0.821 \\
    2011-2014   & 164  & 18 & 24 & 5.22 & 0.815 \\
    2011-2015   & 206  & 14 & 21 & 5.32 & 0.788 \\
    2011-2016   & 231  & 12 & 23 & 5.97 & 0.780 \\
    \hline\hline
  \end{tabularx}
\end{table}

In this table, the number of different clusters for each graph resulted from applying the community detection algorithm of Blondel et al. [2]. By `Mean distance between the nodes' we express the average minimum number of nodes in the graph that lie in between any pair of nodes, as soon as there exists a path linking that pair. This number goes up as the number of nodes in the graph increases but the number of components decreases. Lastly, the `Modularity' column includes the value of the function $Q$ defined in \S3. The strictly decreasing values for $Q$ indicate that as the network evolves, connections between the nodes belonging in a cluster with nodes from different clusters are becoming less sparse.

\bigskip
\textbf{Acknowledgements}.
We wish to express our warmest acknowledgements to our past collaborators Ruth Luo and Jing Mu, with who we explored together clustering algorithms and prepared an initial version of our ``GEAR Collaboration Graph'', as well as to Yannis Drougas for providing his programming expertise and for useful suggestions. We are also thankful to ``ATLAS Web Services'' of the University of Illinois at Urbana-Champaign for continuous web support. The authors acknowledge support from U.S. National Science Foundation grants DMS 1107452, 1107263, 1107367 ``RNMS: GEometric structures And Representation varieties'' (the GEAR Network).

\bigskip
\bigskip
\small{\textsc{Department of Mathematics, University of Illinois at Urbana-Champaign}\\
\,\,1409 W. Green Street Urbana, IL 61801, USA\\
\,\, \emph{E-mail address:} bradlow@illinois.edu}
\bigbreak
\small{\textsc{Oranger Ltd}\\
7-11 Woodcote Road, Wallington, London, SM6 0LH, UK\\
\emph{E-mail address:} i@oranger.me}
\bigbreak
\small{\textsc{Department of Mathematics, University of Illinois at Urbana-Champaign}\\
1409 W. Green Street Urbana, IL 61801, USA\\
\emph{E-mail address:} kydonak2@illinois.edu}
\bigbreak
\small{\textsc{Department of Computer Science, Columbia University}\\
500 West 120 Street, Room 450, MC0401, New York, New York 10027, USA \\
\emph{E-mail address:} xl2601@columbia.edu}
\bigbreak
\small{\textsc{School of Computer Science, Carnegie Mellon University}\\
5000 Forbes Avenue, Pittsburgh, PA 15213, USA\\
\emph{E-mail address:} jiaruix@andrew.cmu.edu}

\end{document}